\documentclass[twocolumn,showpacs,preprintnumbers,amsmath,amssymb]{revtex4}

\usepackage{graphicx}
\usepackage[dvips]{color}
\usepackage{dcolumn}
\usepackage{bm}
\usepackage[nooneline]{subfigure} 
\begin{document}

\preprint{APS/123-QED}

\title{Intrinsic versus Super-rough anomalous scaling in Spontaneous Imbibition}

\author{M. Pradas}
\email{pradas@ecm.ub.es}
\author{A. Hern\'andez-Machado}%
\affiliation{%
Departament d'Estructura i Constituents de la Mat\`eria,\\ Universitat de Barcelona, Avinguda Diagonal 647, E-08028 Barcelona, Spain 
}%
\date{\today}
\begin{abstract}
We study spontaneous imbibition using a phase field model in a two dimensional system with a dichotomic quenched noise. By imposing a constant pressure $\mu_{a}<0$  at the origin, we study the case when the interface advances at low velocities, obtaining the scaling exponents $z=3.0\pm 0.1$, $\alpha=1.50\pm 0.02$ and $\alpha_{loc}= 0.95\pm 0.03$ within the intrinsic anomalous scaling scenario. These results are in quite good agreement with  experimental data recently published. Likewise,  when we increase the interface velocity, the resulting scaling exponents are $z=4.0 \pm 0.1$, $\alpha=1.25\pm 0.02$ and $\alpha_{loc}= 0.95\pm 0.03$. Moreover, we observe that the local properties of the interface change from a super-rough to an intrinsic anomalous description when the contrast between the two values of the dichotomic noise is increased. From a linearized interface equation we can compute analytically the global scaling exponents  which are comparable to the numerical results, introducing some properties of the quenched noise.
\end{abstract}
\maketitle
\section{Introduction}\label{SecI}
The growth of fluctuating interfaces in disordered media has been a subject of much interest  in  last few years\cite{AL04,AN04}. The study of the interface kinetic roughening is a fundamental problem in non-equilibrium statistical physics and has important applications in technological processes and material characterization.  Examples of experimental studies  are fracture surfaces \cite{BO97,LO98}, slow combustion fronts \cite{MA97}, fluid-air interfaces in porous media \cite{SO05,SO02,SO02b,SO03, GE02}. The main common feature of the stable fluctuating interface $h(x,t)$ in such experiments is that it can be described in terms of  scaling laws \cite{BA95}. The basic assumption is to consider the Family-Vicsek hypothesis \cite{FA85}, which supposes a self-affine interface described by a growth exponent $\beta$, a roughness exponent $\alpha$ and a dynamic exponent $z$, with the scaling relation $\alpha=z\beta$. Although this scaling hypothesis is valid in a great variety of cases, different experiments\cite{LO98,SO02b} and numerical models\cite{LE93,LO97} suggest a new scaling description, the so-called anomalous scaling, in the sense that the interface behaves in a not self-affine manner \cite{LO96} and the local and global interface fluctuations are quite different so that a new local roughness exponent $\alpha_{loc}$ has to be introduced.  As pointed out by J.M. Lopez \cite{LO99},  anomalous scaling appears due to the fact that the mean  local slope of the interface  $\langle \nabla h\rangle$ has a nontrivial dynamics. Furthermore, depending on the local exponent values,  anomalous scaling can be classified in a different groups such as the super-rough or intrinsic anomalous  scaling \cite{RA00}.

The advancement of a fluid into a porous medium has received much attention both experimentally and numerically \cite{AL04,AN04}. It is important to remark that experiments can be carried out in two different ways: a) \emph{forced-flow imbibition}, where the invading fluid gets into the porous medium at a constant injection rate and b) \emph{spontaneous imbibition}, where the fluid advances into the medium due to the capillarity. In the present paper we focus on the spontaneous imbibition phenomenon ( see Sec. \ref{SubSecIIA} ), and we base our study on an experimental work  appeared  recently \cite{SO05} where an oil-air interface advances in a horizontal Hele-Shaw cell with a random gap spacing between both glass plates. Traction due to capillarity is so strong that an external negative pressure must be applied  to have a very slow motion of the roughening front. With these experimental conditions, the scaling exponents obtained are $z=3$ , $\alpha\simeq 2$ and $\alpha_{loc}\simeq 1$ within the intrinsic anomalous scaling framework.

 The first attempt to describe spontaneous imbibition numerically was proposed by Dub\'e \emph{et al} \cite{DU99,DU00} employing a phase field model \cite{EL01,HM01,LA05,LU05}. The model is based on the introduction of a partial differential equation (model B in \cite{HH77}) with a quenched noise and it includes the local conservation law of the liquid bulk mass. The resulting scaling exponents in the studied regime are $z=4$, $\alpha=1.25$ and $\alpha_{loc}\simeq 1$ within the super-rough anomalous scaling.

The objective of the present paper is to make a deeper study of the spontaneous imbibition using the phase field model. Our aim is to reproduce the experimental result \cite{SO05} and explain the origin of the intrinsic anomalous scaling. It turns out that the velocity of the interface is a relevant parameter. The exponent $z=3$ indicates a global dynamics driven by capillary forces which occurs at low velocities. Intrinsic anomalous scaling has never numerically been  observed in spontaneous imbibition and a dynamical growth with $z=3$ has only been observed previously in the forced imbibition case \cite{AL04,HM01,LA05}. By increasing the interface velocity we recover the $z=4$ predicted by Dub\'e \emph{et al} \cite{DU99}. In the same line, by increasing the capillary contrast of the quenched noise we can move from a super-rough to an intrinsic anomalous scaling. Furthermore,  the scaling properties of the quenched noise will be discussed to determine analytically the numerical obtained exponents.

The outline of the paper is as follows. In Sec. \ref{SecII} we present both the analytical model to describe spontaneous imbibition and the numerical phase field model. In Sec. \ref{SecIII} we review the statistical properties of rough interfaces describing the anomalous scaling, and in the Sec. \ref{SecIV} we analyze and discuss the obtained numerical results. The final conclusions are given in Sec. \ref{SecV}.
\section{The Model}\label{SecII}
\subsection{The  Spontaneous Imbibition phenomenon}\label{SubSecIIA}
Spontaneous imbibition consists in the advance of a viscous fluid of viscosity $\mu$ (like oil, water) in a porous medium with air, where motion is driven by capillarity. An important classical known result of this kind of motion is that the mean advancing front obeys Washburn's law \cite{WA21}, $H(t)\sim t^ {1/2}$, thus spontaneous imbibition is a case with slowing-down dynamics. 

\begin{figure}[h] 
\centering
\resizebox{0.4\textwidth}{!}{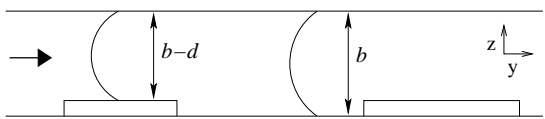}
\caption{Gap variation due to the quenched noise} \label{fig:gapVariation}
\end{figure}

Experimentally, a porous medium can be well reproduced through the so-called Hele-Shaw cell, which  consists on a two horizontal glass plates of size $L_{x}\times L_{y}$,  separated by a small gap $b$ in the $z$-direction, where a fiberglass substrate with copper squares is attached to the bottom plate (see Ref. \cite{SO05}) producing a random gap spacing. A scheme of the meniscus curvature of the interface variation due to the quenched noise is depicted in Fig. \ref{fig:gapVariation}.
\newline
The main effect of this random gap is to produce a random capillarity and permeability. However, in our specific case of spontaneous imbibition, where the motion is controlled solely by capillary forces and the interface advances at very low velocity, the fluctuations due to the permeability will be less important than those produced due to the random capillarity.  Actually, it has been shown that the permeability term depends always on the interface velocity \cite{HM01,LA06,PA03}. In this sense, we can suppose  just the random capillarity as the main destabilizing term.  Note that this approximation could not be valid in the forced-flow imbibition case, where  the interface advances at constant velocity and thus, the random permeability could become an important term, depending on the imposed velocity.

 With this approximation in mind together with the fact that $b<<L_{x}, L_{y}$,  the equation of motion for the viscous fluid
\begin{equation}\label{eq:crreping}
\nabla^{2}\bm{v}=\frac{1}{\mu}\bm{\nabla}p,
\end{equation}
is reduced to the 2-D Darcy's law:
\begin{equation} \label{eq:darcy}
\bar{\bm{v}}(x,y)  = -K\bm{\nabla} p(x,y),
\end{equation}
where $\bar{v}$ represents the averaged velocity over the direction $z$. Since we considerer a constant permeability $K=b^{2}/12\mu$, we can impose the liquid conservation \cite{PA03},  $\bm{\nabla}\cdot\bar{\bm{v}}=0$ , obtaining the Laplace equation for the pressure
\begin{equation}\label{eq:laplaceP}
\nabla^{2}p  = 0.
\end{equation}
The boundary conditions 
 \begin{align}
  \Delta p\vert_{int} & =\sigma \kappa+\eta(x,h), \label{eq:laplaceDisc} {}\\
p(x,y=0) & =P_{a}, \label{eq:AppliedPres}
 \end{align}
complete the equations to describe the spontaneous imbibition phenomenon. Eqns. (\ref{eq:darcy}), (\ref{eq:laplaceP}), (\ref{eq:laplaceDisc}) and (\ref{eq:AppliedPres}) are the equations of the macroscopic model. Eq. (\ref{eq:laplaceDisc}) is the local thermodynamic equilibrium condition at the interface, where  $\sigma$ is the surface tension of the liquid, $\kappa$ is the local curvature of the interface and $\eta(x,h)$ represents the random capillarity due to fluctuations in gap spacing, which it can be approached as $\eta(x,h)\sim 2\sigma\cos\theta/b$. We also assume that the contact angle $\theta$ does not vary. 

On the other hand,  Eq. (\ref{eq:AppliedPres}) is the applied pressure at the origin of the cell and it is the main difference between the spontaneous imbibition and the forced-flow imbibition, since in the last case, a constant velocity is imposed.
\subsubsection*{Washburn's law}
As a simple case of constant capillarity, $\eta(x,y)=\eta_{0}$, the problem is reduced to compute Eqs. (\ref{eq:darcy}) and (\ref{eq:laplaceP})   for the advance of a flat interface with the boundary conditions, Eqs. (\ref{eq:laplaceDisc}) and (\ref{eq:AppliedPres}), where now $\kappa=0$; obtaining the exact solution 
\begin{equation}\label{eq:WashDiferentialEq}
\frac{\textrm{d} H(t)}{\textrm{d} t}=\frac{K(\eta_{0}+P_{a})}{H(t)},
\end{equation}
where $H(t)$ is the position of the interface. Integrating the last expression one obtains the so-called Washburn's law
\begin{equation}\label{eq:WashburnLaw}
H(t)=\sqrt{H_{0}^{2}+2at},
\end{equation}
where $a=K(\eta_{0}+P_{a})$ and we have supposed an initial height of liquid $H_{0}$. Therefore, the expected behavior $H(t)\sim t^{1/2}$ is recovered after an initial transient regime.
\subsubsection*{Linearized Interface Equation }
The general problem of a random field $\eta(x,y)$ is more complex to solve. The interface equation can be derived from Eqs. (\ref{eq:laplaceP}), (\ref{eq:laplaceDisc}) and (\ref{eq:AppliedPres}) using Green function analysis. In this case, the Green function $G(\bm{r},\bm{r'})$ obeys $\nabla^{2}G(\bm{r},\bm{r'})=\delta(\bm{r}-\bm{r'})$ evaluated in the semi-plane $\Omega=\{x,0\leqslant y\}$ with Dirichlet boundary conditions. In order to get an expression for interface fluctuations, the Green equation 
\begin{align}\label{eq:GreenEq}
\int\!\!\!\int_{\Omega_{L}}\!\!\mathnormal{d}\bm{r}'\Big [p(\bm{r}')\nabla'^{2}G(\bm{r},\bm{r}')-G(\bm{r},\bm{r}')\nabla'^{2}p(\bm{r}')\Big ] & = {} \nonumber\\ 
\int_{S_{L}}\!\!\mathnormal{d}\bm{s}'\cdot p(s')\bm{\nabla}'G(s,s')-\int_{S_{L}}\!\!\mathnormal{d}\bm{s}'\cdot G(s,s')\bm{\nabla}'p(s'),
\end{align}
is integrated over the volume of the liquid $\Omega_{L}:\{x,0\leqslant y\leqslant h(x,t)\}$, where $h(x,t)$ is the interface position, and  the Darcy equation, Eq. (\ref{eq:darcy}), is imposed at the interface. Linearising in small deviations of the height $h(x,t)$ around their mean value $H(t)$ in the Fourier space, one can obtain the linearized interface equation 
\begin{equation}\label{eq:LinealH}
\dot{\tilde{h}}_{q}=-\sigma K\vert q\vert q^{2}\tilde{h}_{q}-\dot{H}(t)\vert q\vert \tilde{h}_{q}+K\vert q\vert\tilde{\eta}_{q},
\end{equation}
where we have supposed that correlations grow up slower in time than the position of the averaged interface over $x$, which lies in the condition $\vert q\vert H(t)>>1$. The noise term in Fourier space is taken as $\tilde{\eta}_{q}=\int\mathnormal{d} x \mathnormal{e}^{-iqx}\eta(x,h(x,t))$. The zero mode of the interface position follows the Washburn equation, Eq. (\ref{eq:WashDiferentialEq}), in the form $\dot{H}(t)=a/H(t)$, where now, $a$ depends on the noise average $\langle\eta(x,h(x,t))\rangle_{x}$ instead of $\eta_{0}$.

The linearized equation, Eq. (\ref{eq:LinealH}), is the same equation obtained in Ref. \cite{DU99} and contains two mechanisms to damp the interface fluctuations produced by the quenched noise $\tilde{\eta}_{q}$. At small scales, the surface tension $\sigma$ controls the fluctuations introducing a dynamic exponent $z=3$. Actually, it is the regime observed experimentally in Ref. \cite{SO05}, indicating that a linear description is valid in that situation. The long scales are damped by the advancement of the front $\dot{H}(t)$. These two effects are separated by a dynamic crossover length:
\begin{equation}\label{eq:CrossoverL}
\xi_{\times}=2\pi\Big( \frac{\sigma K}{\dot{H}(t)}\Big)^{1/2}.
\end{equation}
Since the mean height of the interface follows the Washburn law, this crossover length increases with time, $\xi_{\times}\simeq t^{1/4}$. The numerical results presented in Ref. \cite{DU99,DU00} shows that this crossover length acts as \emph{cut off} for the fluctuations growth due to the interface is always asymptotically flat on length scales larger than $\xi_{\times}$. Thus if the velocity of the interface is large enough, the roughening of the interface is controlled by $\xi_{\times}$ which gives rise to a dynamic exponent $z=4$.
\subsection{The Phase Field Model}\label{SubSecIIB}
Our numerical approach is based on the phase field model, already introduced in Ref. \cite{DU99}, where a conserved field $\phi$ is used to represent the two phases of the problem: liquid/air with the equilibrium values $\phi_{eq}= +1/-1$ respectively. The dynamics of the field is controlled by a continuity equation based on a time-dependent Ginzburg-Landau model with conserved order parameter (model B in \cite{HH77} ) $\partial\phi/\partial t=\nabla M\nabla\mu$ where $\mu=\delta \mathcal{F}/\delta\phi$ is the chemical potential and the free energy is taken with the form $\mathcal{F}[\phi]=\int\mathnormal{d}\pmb{r}(V(\phi)+(\epsilon\nabla\phi)^{2}/2)$. A double well potential is also chosen with a linear random term, which destabilizes one of the phases and forces  the interface to advance:
\begin{equation}\label{eq:DoubleWell}
V(\phi)=-\frac{1}{2}\phi^{2}+\frac{1}{4}\phi^{4}-\eta(\pmb{r})\phi.
\end{equation}
Here, the noise $\eta(\pmb{r})$ plays the capillarity role. In our numerical integrations we use a dichotomic noise distributed in a 2-D system with the values:
\begin{equation}\label{eq:Noise}
\eta = \left\{\begin{array}{ll}
     \eta_{0} & \\
      & \\
     \frac{\eta_{0}}{1-\eta_{A}} &
                \end{array} \right. {}\\
\end{equation}
The noise is imposed in the same way that in the experimental work in a Hele-Shaw cell \cite{SO05}, and it is distributed in such way that the sites with $\eta_{0}$ occupies the $65\%$ of the whole system. Even though in such experiments capillarity arises as a 3-D effect, we can relate our noise parameters with the Hele-Shaw gap spacing $b$ as $\eta_{0}\sim 1/b$ and $\eta_{A}\sim d/b$. So the $\eta_{A}$ parameter controls the contrast between both noise values whereas the $\eta_{0}$ corresponds to the capillarity of the cell with a fixed gap.

The resulting equation for the phase field reads
\begin{align}\label{eq:Phase-Field}
\frac{\partial\phi}{\partial t} & =  \bm{\nabla} M\bm{\nabla}\big[V'(\phi)-\epsilon^{2}\nabla^{2}\phi\big] \nonumber \\ 
{} &=\bm{\nabla} M\bm{\nabla}\big[-\phi+\phi^{3}-\epsilon^{2}\nabla^{2}\phi-\eta(\bm{r})\big],
\end{align}
where $M$ is a parameter which we take  constant  at the liquid phase ($\phi>0$) and zero at the air phase ($\phi<0$).\\ 
In order to recover  the macroscopic results described by Eqs. (\ref{eq:darcy}), (\ref{eq:laplaceP}), (\ref{eq:laplaceDisc}) and (\ref{eq:AppliedPres}), a matched asymptotic expansion in the \emph{Sharp Interface Limit}, $\epsilon\to 0$, is necessary. This expansion has been done in Ref. \cite{HM03}.  Although in that case the phase field is used to describe the Saffman-Taylor problem, a similar  expansion is also valid in our case of a stable interface, with the identification of the phase field parameters as: 
\begin{equation} \label{eq:Parameters}
p= \phi_{eq}\mu_{1},  \quad
K= \frac{M}{2\phi_{eq}^{2}}, \quad
\sigma =\frac{1}{2}\int\mathrm{d} w \Big(\frac{\partial \phi_{0}}{\partial w}\Big)^{2},
\end{equation}
where $\mu_{1}$ is the first order in $\epsilon$ term of the chemical potential in the asymptotic expansion and $\phi_{0}=-\phi_{eq}tanh(w/\sqrt{2})$ is a kink solution of the phase field equation for a flat interface in equilibrium, and it corresponds to the zero order in $\epsilon$ term of the order parameter. The variable $w$ is an inner coordinate of the interface, introduced in the expansion, and it is perpendicular to it at any point.

\begin{figure}[h]
\centering
\resizebox{0.35\textwidth}{!}{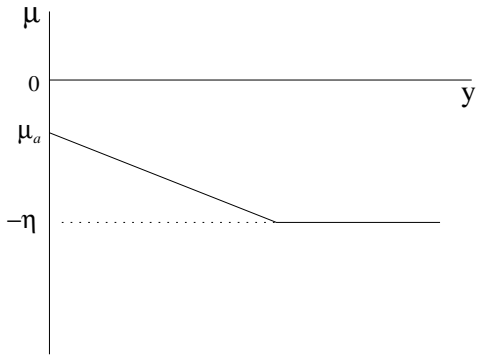}
\caption{Variation of the chemical potential in an arbitrary time along the $y$-direction. The velocity of the interface is related to $-\partial_{y}\mu$.} \label{fig:Mu}
\end{figure}

Since the chemical potential is equivalent to the pressure, it has to be fixed at the origin of our system  to reproduce  spontaneous imbibition, $\mu(x,y=0)=\mu_{a}$. In Fig. \ref{fig:Mu} there is plotted a scheme of the chemical potential variation along the $y$-direction. The imbalance created in the chemical potential by $\mu_{a}+\eta$ causes the interface to advance following the Washburn law, Eq. (\ref{eq:WashburnLaw}).

Throughout the paper we call applied pressure to the value $\mu_{a}$. We will also suppose an initial height of liquid $H_{0}$ to get a low initial velocity, Eq. (\ref{eq:WashDiferentialEq}).
\section{Scaling of rough interfaces}\label{SecIII}
The statistical properties of a one-dimensional interface defined by a function $h(x,t)$ are usually described in terms of the fluctuations of $h$. The Family-Vicsek  scaling hypothesis \cite{FA85} ensures the existence of a lateral correlation length $\ell_{c}$ expected to increase as $\ell_{c}\sim t^{1/z}$ until it reaches the system size $\ell_{c}=L$, which it defines a saturation time $t_{s}\sim L^{z}$. On the other hand, the vertical correlation length or interface \emph{global} width, defined as  
$W(L,t)=\langle \overline{[h(x,t)-\overline{h}]^{2}}\rangle^{1/2}$, increases as $W(L,t)\sim t^{\beta}$ for $t<t_{s}$ 
and becomes constant $W(L,t)\sim L^{\alpha}$ for $t \geqslant t_{s}$. Here, $\langle ..\rangle$ denotes average over different noise realizations and the overbar is an spatial average in the $x$ direction. Alternatively, one may study the correlations over a distance $\ell \ll L$ through the \emph{local} width, 
$w(\ell,t)=\langle \langle{[h(x,t)-\langle{h}\rangle_{\ell}]^{2}\rangle_{\ell}}\rangle^{1/2}$, 
where $\langle...\rangle_{\ell}$ denotes an average over $x$ in windows of size $\ell$. Since in the Family-Vicsek scenario the local scales grows in the same way as the global ones, the scaling behavior of the surface can be obtained by looking at the local width.

Another useful quantity also used to obtain the roughness exponent $\alpha$ is the power spectrum of the interface $S(k,t)=\langle\tilde{h}_{k}(t)\tilde{h}_{-k}(t)\rangle$, where $\tilde{h}_{k}(t)$ is the interface Fourier transform. In the  Family-Vicsek assumption it scales as
\begin{equation} \label{ScalingAnsatzFV}
S(k,t)=k^{-(2\alpha+1)}s_{FV}(kt^{1/z}),
\end{equation}
where $s_{FV}$ is the scaling function 
\begin{equation} \label{eq:ScalingFunctionFV}
 s_{FV}(u) \sim  \left\{\begin{array}{ll}
     \textrm{const.} & \textrm{if} \quad u \gg 1 \\
     u^{2\alpha+1}  &  \textrm{if} \quad u \ll 1.\\
       \end{array} \right. 
\end{equation}
However, experimental results and several growth models have appeared in the last decade showing that global and local scales are not equivalent. It is the so-called \emph{anomalous scaling} \cite{LO97}. In this sense, a new scaling \emph{ansatz} is needed to characterize both global and local growth.

This new scaling \emph{ansatz} was proposed by  Ramasco \emph{et al} \cite{RA00} and it was a generalization of the Family-Vicseck scaling function, Eq. (\ref{eq:ScalingFunctionFV}),
\begin{equation} \label{eq:ScalingSPCAnsatz}
S(k,t)=k^{-(2\alpha+1)}s_{A}(kt^{1/z}),
\end{equation}
where now the scaling function has the general form
\begin{equation} \label{eq:ScalingFunctionSPC}
 s_{A}(u) \sim  \left\{\begin{array}{ll}
     u^{2(\alpha-\alpha_{s})} & \textrm{if} \quad u \gg 1 \\
     u^{2\alpha+1}  &  \textrm{if} \quad u \ll 1,\\
       \end{array} \right. 
\end{equation}
being $\alpha_{s}$  the spectral roughness exponent. This scaling behavior of the power spectrum leads to the following dynamic scaling for the local width:
\begin{equation} \label{eq:ScalingWAnsatz}
w(\ell,t)= t^{\beta}g(\ell / t^{1/z}),
\end{equation}
with the corresponding scaling function 
\begin{equation} \label{eq:ScalingFunctionW}
 g(u) \sim  \left\{\begin{array}{ll}
     u^{\alpha_{loc}} & \textrm{if} \quad u \ll 1 \\
     \mathrm{const.}  &  \textrm{if} \quad u \gg 1,\\
       \end{array} \right. 
\end{equation}
where $\alpha_{loc}$ is the  local rough exponent and it characterizes the roughness at small scales. One of the implications of the anomalous scaling is that the local width saturates when the system size saturates as well, i.e, at times $t_{s}$ and not at the local time $t_{\ell}\sim \ell^{z}$ as occurs in the Family-Vicsek scaling. There is an intermediate regime between $t_{\ell}$ and $t_{s}$ where the local width grows as $w(\ell,t)\sim t^{\beta^{*}}$ with $\beta^{*}=\beta-\alpha_{loc}/z$. 

Different types of scaling arise from the general scaling function, Eq. (\ref{eq:ScalingSPCAnsatz}). For $\alpha_{s}<1$ it is always accomplished that $\alpha_{loc}=\alpha_{s}$; in this case the Family-Vicsek scaling is achieved when $\alpha_{loc}=\alpha$, and the intrinsic anomalous  scaling appears when $\alpha>\alpha_{loc}$. The main feature of such an anomalous scaling is the existence of a temporal shift in the power spectrum due to the difference between $\alpha_{loc}$ and $\alpha$. On the other hand, for $\alpha_{s}>1$ always occurs that $\alpha_{loc}=1$ and the super-rough anomalous scaling appears when $\alpha=\alpha_{s}$. 
\section{Numerical Results}\label{SecIV}
The numerical study of the imbibition phenomenon is made by means of the integration of the phase field equation, Eq. (\ref{eq:Phase-Field}), in a rectangular lattice of $\Delta x =1$ grid space, with periodic boundary conditions in the $x$ direction. We have also considered a smaller grid space $\Delta x=0.5$ and we reproduce identical results as we explain below. The interface position $h(x,t)$ is approached by a linear interpolation of the zero of the phase field,  $\phi(x,h(x,t);t)=0$. In all of our numerical integrations we will suppose an initial height $H_{0}=20$ and the noise value $\eta_{0}=3.0$.  Our study is focused to obtain the different scaling exponents and its dependence on the applied pressure, $\mu_{a}$,  and on the contrast between noise values $\eta_{A}$.

\subsection{Variation in the applied pressure}\label{SubSecIVA}
\begin{figure}[!]%
\centering
\includegraphics[width=0.45\textwidth, origin=c]{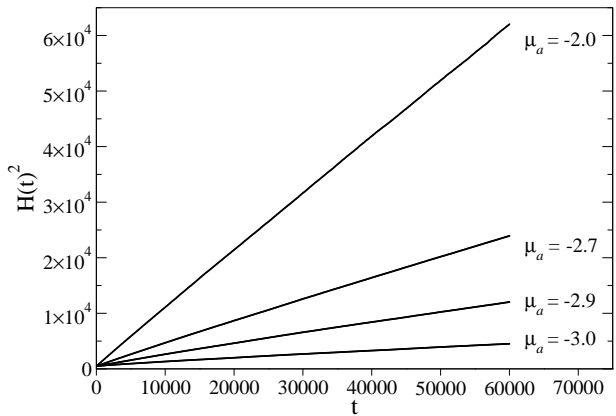}%
\caption{Squared averaged interface height position,  as function of time, follows Washburn law for different applied pressure $\mu_{a}$.\\}\label{fig:Wahburn}
\end{figure}

\begin{figure}[!]%
\vspace{0.4cm}
\centering
\includegraphics[width=0.5\textwidth, origin=c]{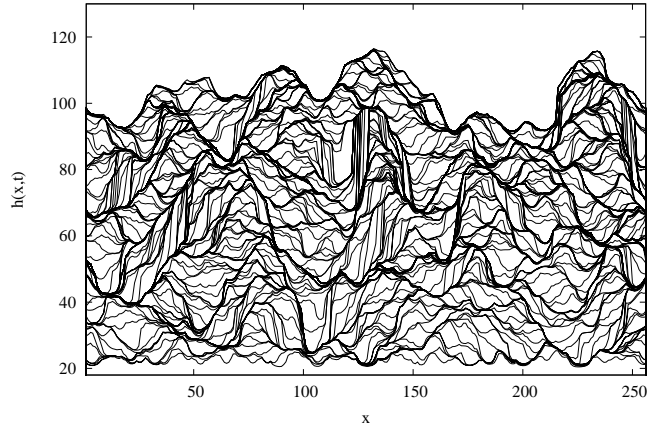}%
\caption{Temporal evolution of the interface at equal time intervals. Numerical parameters are $\eta_{0}=3.0$, $\eta_{A}=0.3$ and $\mu_{a}=-3.0$ }\label{fig:IntZ3}
\end{figure}

Different applied pressures at the origin have been studied. In all of them the Washburn law, Eq. (\ref{eq:WashburnLaw}), is well accomplished as we can see from Fig. \ref{fig:Wahburn}.
 As a first case to analyze, we choose the numerical parameters  $\eta_{A}=0.3$ and $\mu_{a}=-3.0$. In Fig. \ref{fig:IntZ3} there is plotted the temporal evolution of the interface. Note that the applied pressure has the same value in module than the noise value $\eta_{0}$ and thus there are grid sites where the interface is locally pinned (Fig. \ref{fig:Mu}). In this situation, the interface has a slow motion and advances in avalanches.

\begin{figure}[!]%
 \centering %
 \subfiguretopcapfalse
 \subfigure{{\includegraphics[width=0.45\textwidth,origin=c,angle=0]{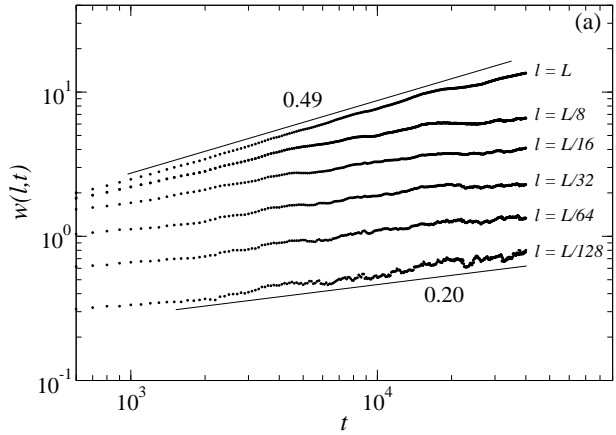}}}\\%
 \vspace*{0.6cm}%
 \subfigure{\includegraphics[width=0.45\textwidth,origin=c,angle=0]{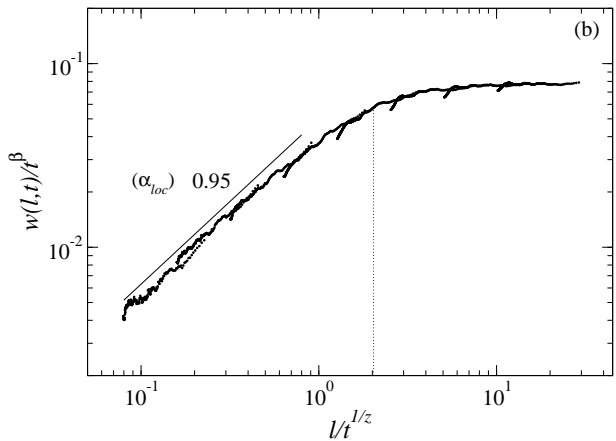}}\\%
 \vspace*{0.6cm}%
 \subfigure{\includegraphics[width=0.45\textwidth,origin=c,angle=0]{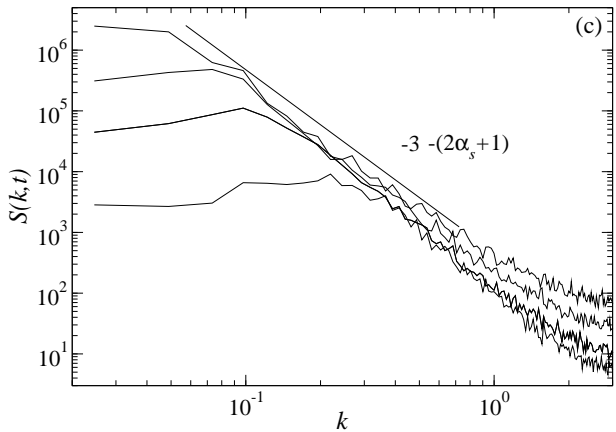}}%
 \caption{Statistical description of interfaces when $\mu_{a}=-3.0$ and $\eta_{A}=3.0$: (a) Log-log plot of the interfacial width as function of time in different windows sizes $\ell$. The data fit shows a growth exponent $\beta=0.49\pm 0.03$ and the local one $\beta^{*}=0.20\pm 0.04$. (b) Collapse of the interface local width function, $w(\ell,t)$, which is scaled by $t^{\beta}$ versus $\ell /t^{\beta/\alpha}$, where $\alpha=1.50\pm 0.02$, and then $z=3.0\pm 0.1$  follows from $z=\alpha/\beta$. The picture also shows a local roughness exponent $\alpha_{loc}= 0.95 \pm 0.03$. (c) Power spectra of the interface $h(x,t)$ at different times. It shows that $\alpha_{s}=1.0\pm 0.1$.}\label{fig:StatisticalZ3}
\end{figure}

Using the scaling concepts described in the last section, we do a statistical description of the interface fluctuations, Fig. \ref{fig:StatisticalZ3}.
From the interface local width we obtain the global growth exponent $\beta=0.49\pm 0.03$ and the local one $\beta^{*}=0.20 \pm 0.04$, clearly different from $0$, indicating the existence of anomalous scaling. On the other hand, from the interface power spectrum we obtain the spectral roughness exponent $\alpha_{s}=1.0\pm 0.1$. The temporal shift observed in the power spectrum indicates that $\alpha>\alpha_{s}$ and since $\alpha_{s}$ is not greater than $1$, the relation $\alpha_{loc}=\alpha_{s}$ should be accomplished \cite{RA00}. Therefore, the global roughness exponent $\alpha$ can not be achieved directly from the power spectrum but it has to be obtained by means of a collapse of the interface local width function, Eq. (\ref{eq:ScalingWAnsatz}). The result of the best collapse is shown in Fig. \ref{fig:StatisticalZ3}b from which we obtain that  $\alpha=1.50 \pm 0.02$ and $z=3.0\pm 0.1$. This plot also shows a local roughness exponent of $\alpha_{loc}=0.95\pm0.03$, which is very close to the value of the roughness exponent, $\alpha_{s}$. These set of scaling exponents are compatible within the intrinsic anomalous scaling scenario, and they are clearly different from those obtained in Ref. \cite{DU99}. 

\begin{figure}[!]%
\centering
\includegraphics[width=0.45\textwidth, origin=c]{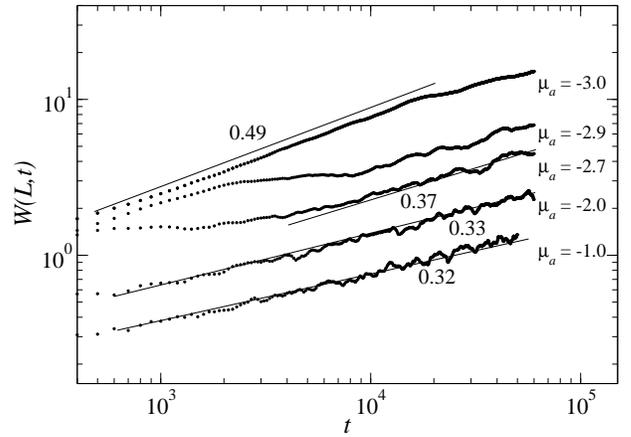}%
\caption{ Computed global width $W(L,t)$ by changing the applied pressure $\mu_{a}$. The lines show a fit data which allows us to obtain the $\beta$ exponent}\label{fig:Global}%
\end{figure}

Let us now increase the interface velocity by going up the applied pressure, $\mu_{a}$. Following the same statistical procedure as before, we compute the whole set of scaling exponents for different values of $\mu_{a}>-3.0$. Fig. \ref{fig:Global} shows the global width $W(L,t)$ in each case.
\begin{figure}[!]%
 \centering %
 \subfiguretopcapfalse
 \subfigure{{\includegraphics[width=0.45\textwidth,origin=c,angle=0]{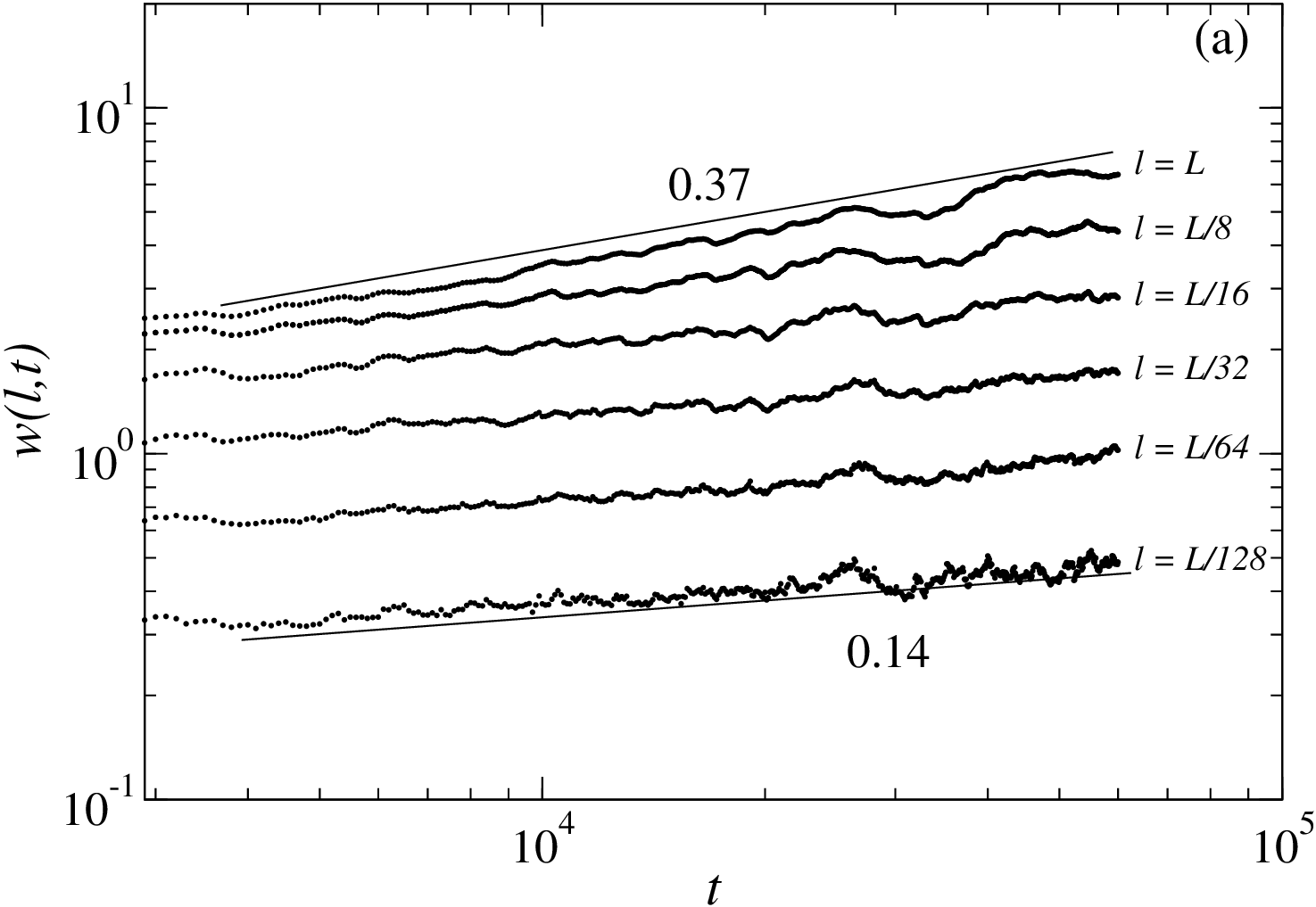}}}\\%
 \vspace*{0.6cm}%
 \subfigure{\includegraphics[width=0.45\textwidth,origin=c,angle=0]{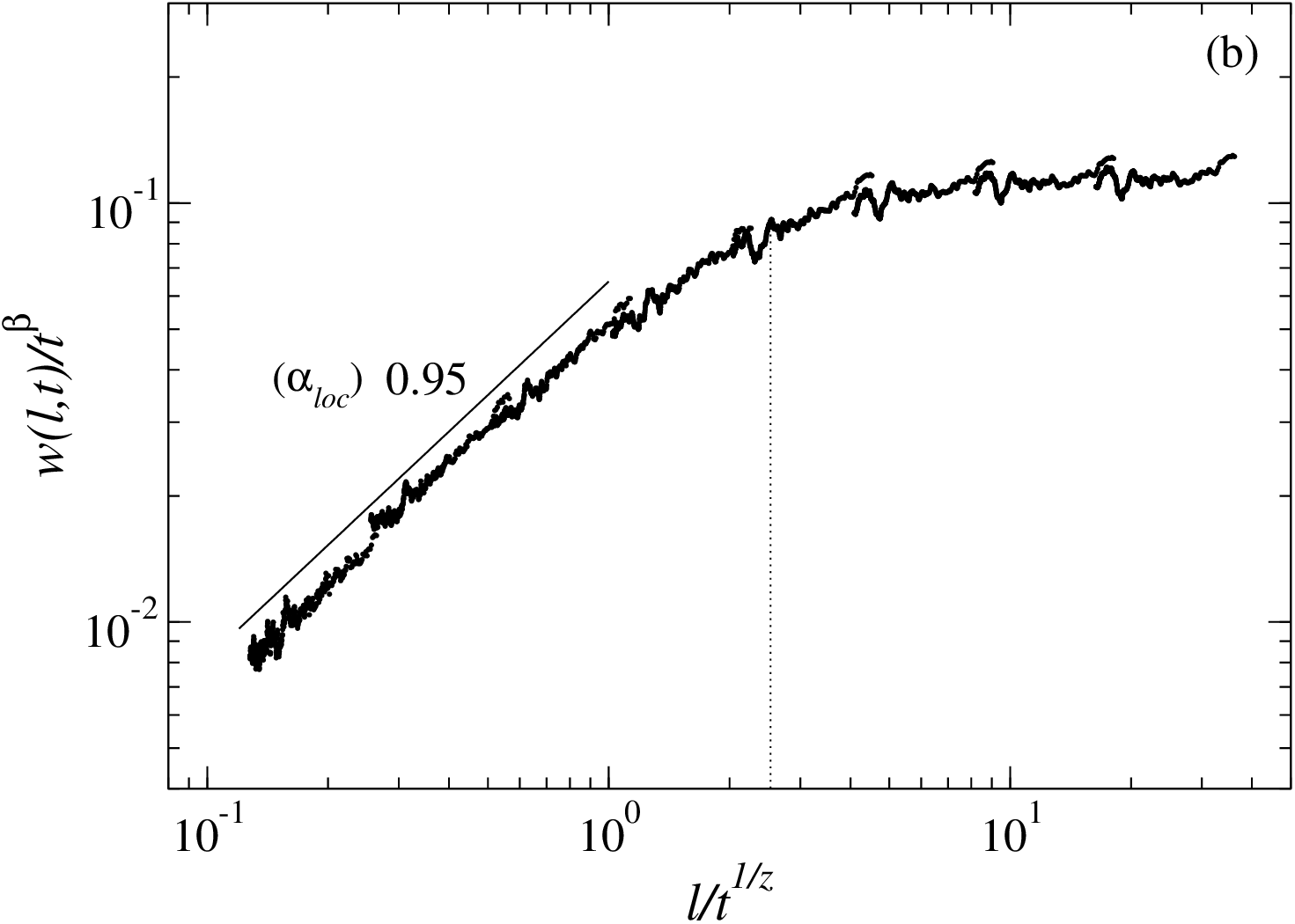}}\\%
 \vspace*{0.6cm}%
 \subfigure{\includegraphics[width=0.45\textwidth,origin=c,angle=0]{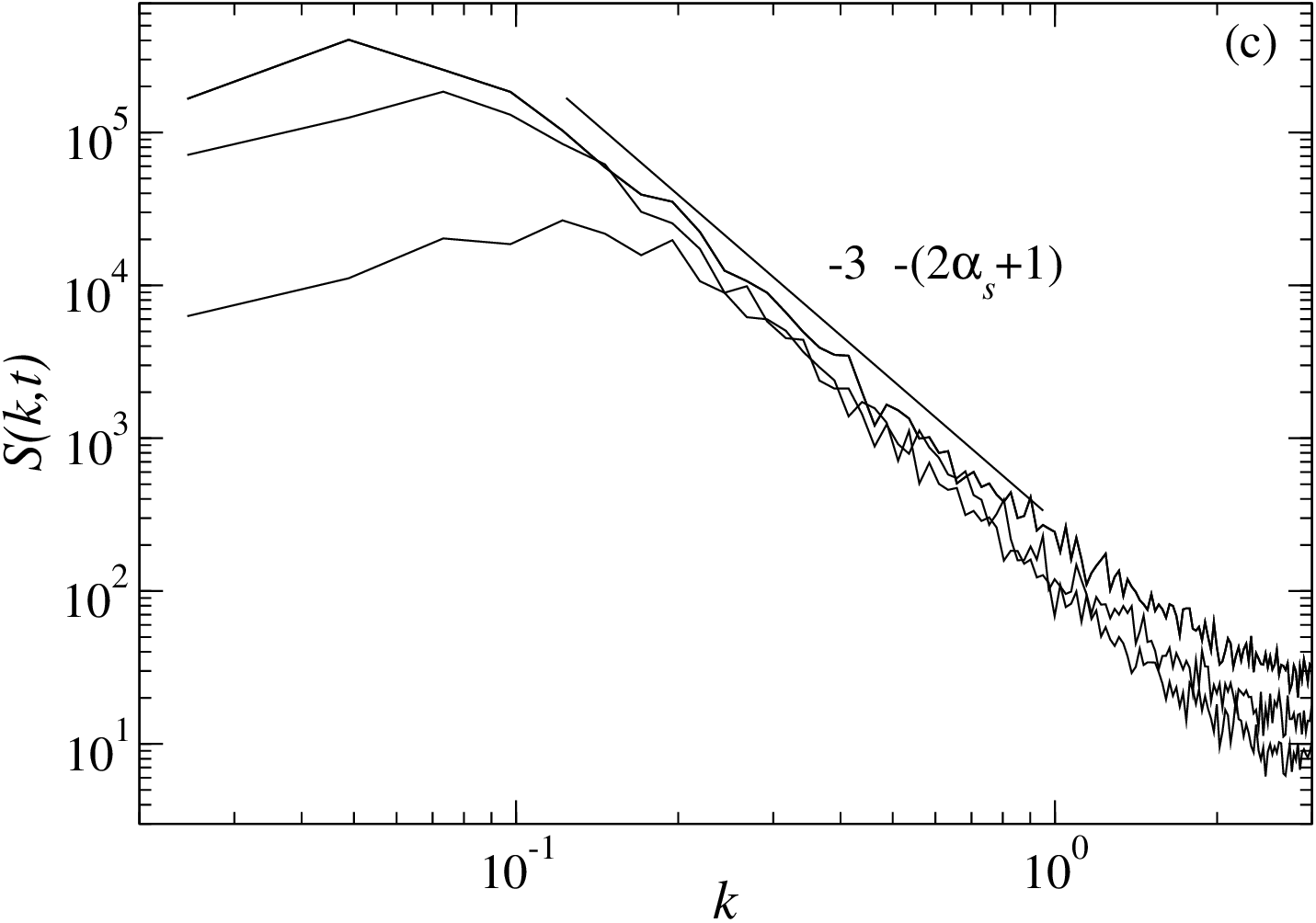}}%
 \caption{Statistical description of interfaces when $\mu_{a}=-2.7$ and $\eta_{A}=3.0$: (a) Log-log plot of the interfacial width as function of time in different windows sizes $\ell$. The data fit shows a growth exponent $\beta= 0.37\pm 0.03$ and the local one $\beta^{*}= 0.14 \pm 0.04$.(b) Collapse of the interface local width function, $w(\ell,t)$, which is scaled by $t^{\beta}$ versus $\ell /t^{\beta/\alpha}$, where $\alpha=1.50\pm 0.02$, and then $z=4.0\pm 0.1$  follows from $z=\alpha/\beta$. The picture also shows a local roughness exponent $\alpha_{loc}= 0.95 \pm 0.03$. (c) Power spectra of the interface $h(x,t)$ at different times. It shows that $\alpha_{s}=1.0\pm 0.1$.}\label{fig:StatisticalZ4}
\end{figure}%
We observe a crossover from $\beta\simeq 0.5$ ($\mu_{a}=-3.0$) to $\beta\simeq 0.37$  ($\mu_{a}=-2.7$).  In the specific case of $\mu_{a}=-2.7$, Fig. \ref{fig:StatisticalZ4}, the different scaling exponents  are $\alpha=1.5\pm 0.02$, $z=4.0\pm0.1$, $\beta= 0.37\pm0.03$, and $\alpha_{loc}= 0.95\pm0.03$. 

Finally, the results  obtained in Ref. \cite{DU99},  $\alpha=1.25$, $z=4$, $\beta\simeq 0.32$ and $\alpha_{loc}\simeq 1$, are recovered when we increase the applied pressure up to $\mu_{a}=-1.0$. So it is important to remark that decreasing the applied pressure produces three important effects: 1) the roughness exponent $\alpha$ changes from $1.25$ to $1.5$, 2) the interface local growth change from a super-rough to an intrinsic anomalous scaling description and 3) the dynamic exponent $z=3$ obtained in experimental work is attained when the interface advances at  low velocities. In Table \ref{tab:exp} we can see the different scaling exponents computed for different values of $\mu_{a}$. 

The meaning of large and low velocities has to be understood as follows. The interface advances at low velocity when the pinning effects  become important and the crossover length is large enough to observe an initial regime described by $z=3$. Therefore, in this sense, we call low velocities  the cases of $\mu_{a}<-2.0$ and large velocities  the other ones, $\mu_{a}\ge-2.0$.

Concerning the low velocities regime, there is some numerical inaccuracy in the determination of the interface position due to the slow advance of the interface. In order to check our numerical results, we have performed the case of $\mu_{a}=-3.0$ with a smaller grid space. We have used a $\Delta x=0.5$ and no difference between the case of $\Delta x=1.0$ has been observed.

\begin{table}[!]%
\centering%
\begin{ruledtabular}
\begin{tabular}{c|cccccc}%
  $\mu_{a}$ &  $\alpha$  &  $z$  & $\beta$ & $\beta^{*}$   &  $\alpha_{s}$  &  Scaling\\ \hline%
  -1.0   &     1.25   &    4   &   0.32  &    0.10  &    1.25    &   S-R  \\
  -2.0   &     1.34   &    4   &   0.33  &    0.12  &     0.9     &   I-A   \\
  -2.7   &     1.5     &    4   &   0.37  &    0.14  &     0.9     &   I-A   \\
  -3.0   &     1.5     &    3   &   0.49  &    0.20  &     0.9     &   I-A   \\
 \end{tabular}%
\end{ruledtabular}
\caption{Scaling exponents obtained when the applied pressure $\mu_{a}$ decreases. The contrast of the noise values is kept constant $\eta_{A}=0.3$.}\label{tab:exp}%
\end{table}%
\subsubsection*{The crossover and correlation length}
\begin{figure}[!]%
 \centering %
 \includegraphics[width=0.4\textwidth,origin=c,angle=0]{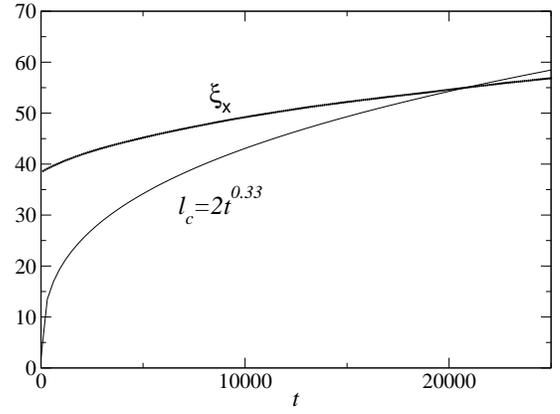}
 \caption{Estimation of the crossover and correlation length when the dynamic exponent $z=3$ is observed.}\label{fig:Cross}
\end{figure}

So far we have seen that an initial regime, dominated by the surface tension term of  Eq. (\ref{eq:LinealH}), can be observed if the interface velocity is low enough. The point is that the correlation length $\ell_{c} =A\cdot t^{1/z}$ is smaller than the crossover length $\xi_{\times}$ during the time interval studied. In order to check it, we are interested in estimating the crossover and correlation length. First, we suppose that in the case of $\mu_{a}=-2.7$, in the regime of $z=4$,  the crossover length, $\xi_{\times}=A\cdot t^{1/4}$, acts effectively as the correlation length as Dub\'e \emph{et al} have pointed out \cite{DU99}, where the constant $A$ can be approached from the inflection point of the width collapse, Fig. \ref{fig:StatisticalZ4}b, being $A\simeq 2.5$. On the other hand we know that the crossover length has the expression $\xi_{\times}=C(2/a)^{1/4}\cdot t^{1/4}$, obtained from Eq. (\ref{eq:CrossoverL}) in the limit $H_{0}^{2}<2at$, where $a$ follows from Washburn law, Eq. (\ref{eq:WashburnLaw}). With these two expression we can obtain the constant $C\simeq 1.4$,  which allow us to evaluate the crossover length in its full expression, Eq.  (\ref{eq:CrossoverL}), when $\mu_{a}=-3.0$. In such case the correlation length grows up as $\ell_{c}=2\cdot t^{1/3}$, where now the parameter $2$ has been estimated from the width collapse in Fig. \ref{fig:StatisticalZ3}b. These results are depicted in Fig. \ref{fig:Cross}, where we can observe that during the studied time interval, the condition $\xi_{\times}>\ell_{c}$ is accomplished, and thus the obtained dynamical exponent $z=3$ is fully justified.

It should be noted from Eq.(\ref{eq:WashburnLaw}) that the crossover length takes a certain time $\tau=H_{0}^{2}/2a$ to reach its asymptotic form $\xi_{\times}\simeq t^{1/4}$. It could produce some effect on the interface scaling as long as $t_{\times}\leq\tau$, where $t_{\times}$ is defined as the time when $\ell_{c}=\xi_{\times}$. As we can see in Fig. \ref{fig:Global}, in both cases of $\mu_{a}=-2.9$ and $\mu_{a}=-2.7$, there is a regime, during the transition from $z=3$ to $z=4$, where the interface  global width does not increase. Computing the parameter $a$ from a fit of the Washburn law, it turns out that $\tau\sim 2\cdot 10^{3}$ and $10^{3}$ for $\mu_{a}=-2.9$ and $-2.7$  respectively, indicating that $\tau\sim t_{\times}$ in both cases. For larger values of the applied pressure this effect occurs in an initial stage of the interface growth and it is not observed.

An interesting point to comment from these numerical results is that the roughness exponent $\alpha$ does not change its value when the correlation length  catch up the crossover length, that is, the growth exponent $\beta$ varies accordingly with the change of the dynamic exponent from $z=3$ to $z=4$ keeping constant the $\alpha$ value (see Fig. \ref{fig:Global}). Therefore, it lead us to think that \emph{the crossover length does not change the scaling properties of the interface except for the dynamic exponent $z$}. It can be very useful to extract information from the linear Eq. (\ref{eq:LinealH}). Since the initial  interface velocity has a finite value greater than zero,  the condition $t_{\times}>0$ is always accomplished and therefore there will be always an initial regime where fluctuations are dominated by the surface tension. Note that this initial regime could be very short for a large initial  velocity of the interface. Hence, we can suppose that the scaling exponents $\alpha$ and $\alpha_{loc}$, observed when the crossover length  is controlling the fluctuations growth, are the same than those corresponding to the initial regime described by $z=3$.
\subsubsection*{Intrinsic anomalous scaling in the low velocity regime}
Our numerical results show the influence of the applied pressure on the scaling exponents. We have seen that the intrinsic anomalous scaling appears when we decrease the applied pressure. Actually, the large local slopes, which appear when the interface velocity is decreased, cause that the scaling properties of the interface follows an intrinsic anomalous description.

In order to approach analytically the global exponents in the case of low velocities we should know the scaling properties of the noise. Likewise, the fact the interface is locally pinned may introduce some difficulties on the quenched noise and we can not know \emph{a priori} how the noise scales. However, the roughness exponent $\alpha=1.5$ seems to indicate that the interface feels an effective white columnar noise, that is $\eta(x)$ with correlations $\langle\eta(x)\eta(x')\rangle\simeq\delta(x-x')$. Indeed, if we rescale the linearized interface equation, Eq. (\ref{eq:LinealH}), by
\begin{equation}\label{eq:rescaling}
x\to bx \quad h\to b^{\alpha}h \quad t\to b^{z}t,
\end{equation}
the noise scales then as
\begin{equation}\label{eq:ScaleColumnarNoise}
\eta(x)\to  b^{-\frac12}\eta(x),
\end{equation}
and the roughness exponent we obtain is $\alpha=1.5$. This roughness exponent is the same value obtained in the local QEW model with columnar noise \cite{LO95}.

It should be pointed out that this assumption of columnar noise is related to the number of sites where the interface is pinned; note that the fractions of the interface which are locally pinned increases as the applied pressure decreases up to values lower than $\mu_{a}<-3.0$ where the interface does not advance at all. So in this sense, the roughness exponent $\alpha=1.5$ is observed when the interface advances near a critical situation of pinning. These results agree with the hypothesis of Parisi \cite{PA92} who suggest that near the pinning transition, both QEW and QKPZ models follows the same universality class than in the QEW model with a columnar noise $\eta(x)$. However, we have to remark that only the roughness exponent $\alpha=1.5$ obtained in our model adjusts with such an universality class.

On the other hand, when the applied pressure is increased and therefore we move further away from the critical situation we recover the results of $\alpha=1.25$ corresponding to the university class of the local QEW model with quenched noise $\eta(x,h)$\cite{AL04}. In this sense, the other observed roughness exponent $\alpha\simeq 1.34$ can be understood as an intermediate regime between the value $\alpha=1.25$ and $\alpha=1.5$ when the applied pressure is decreased. Likewise, in the next section we will discuss an analytical approach which describes the quenched noise scaling in the high velocity regime, when the values $\alpha=1.25$ and $\alpha=1.33$ are observed.

\begin{figure}[!]
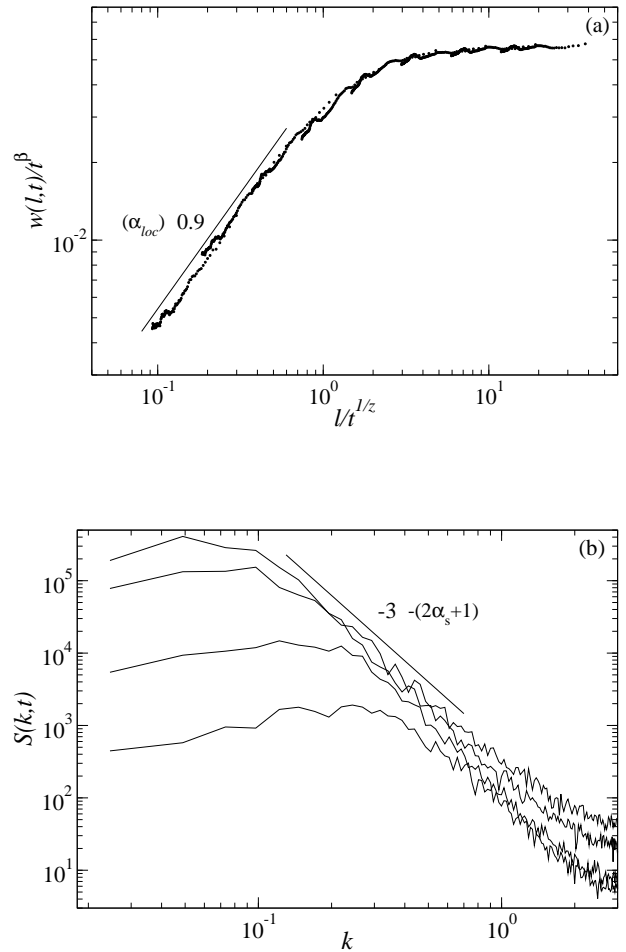
%
 \centering %
 \subfiguretopcapfalse
 \subfigure{{\includegraphics[width=0.45\textwidth,origin=c,angle=0]{colG.eps}}}\\%
 \vspace*{0.6cm}%
 \subfigure{\includegraphics[width=0.45\textwidth,origin=c,angle=0]{SPCg.eps}}\\%
 \caption{Statistical description of interfaces described by Eq. (\ref{eq:Phase-Field}) with a gaussian noise distributed at the $100$\% of the system with mean $\langle\eta\rangle=3.0$ and strength $\Delta\eta=0.6$. The applied pressure is $\mu_{a}=-2.5$.(a) Collapse of the interface local width function, $w(\ell,t)$, which is scaled by $t^{\beta}$ versus $\ell /t^{\beta/\alpha}$, where $\alpha=1.51\pm 0.02$, $\beta=0.51\pm0.03$ and then $z=3.0\pm 0.1$  follows from $z=\alpha/\beta$. The picture also shows a local roughness exponent $\alpha_{loc}= 0.9 \pm 0.03$. (c) Power spectra of the interface $h(x,t)$ at different times. It shows that $\alpha_{s}=1.0\pm 0.1$.}\label{fig:StatisticalZ3g}
\end{figure}

In order to check the validity of this numerical results, we have performed the simulations with another kind of noise. We have used a gaussian field distributed at the $100$\% of the system with mean $\langle\eta\rangle=3.0$ and strength $\Delta\eta=0.6$. When the applied pressure is $\mu_{a}=-2.5$, the interface advance at low velocities and we recover the same results of $\alpha=1.5$ and $z=3$ within the intrinsic anomalous scaling, Fig \ref{fig:StatisticalZ3g}. Therefore, the columnar noise assumption is also correct when we use a different noise distribution.
\begin{figure*}[!]
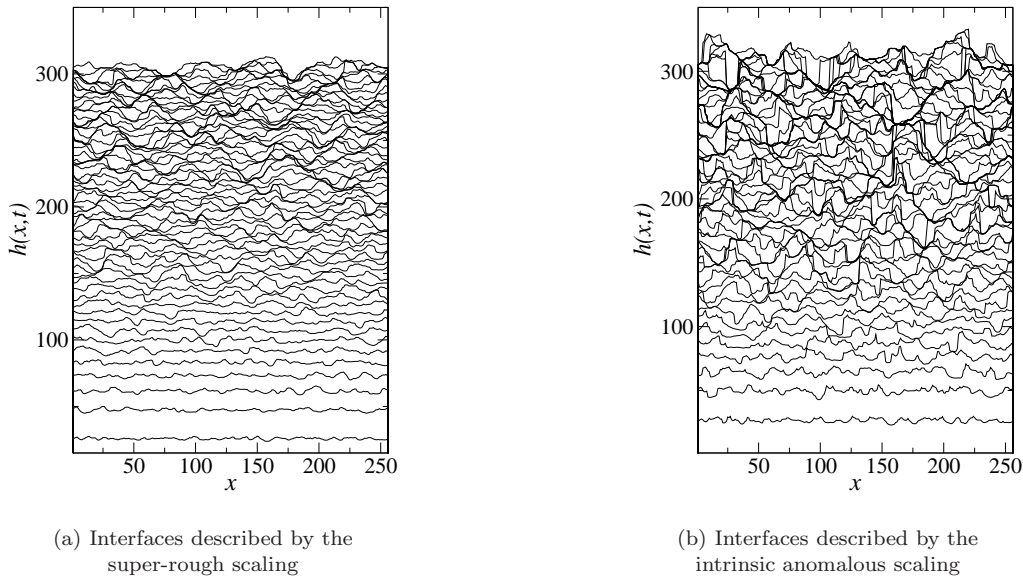
%
\centering
\subfigure[Interfaces described by the super-rough  scaling\label{fig:SRInt}]{{\includegraphics[width=0.29\textwidth,origin=c,angle=0]{intZ4.eps}}}%
\hspace{3.0cm}
\subfigure[Interfaces described by the intrinsic anomalous scaling\label{fig:IAInt}]{{\includegraphics[width=0.29\textwidth,origin=c,angle=0]{intAZ4.eps}}}\\%
\caption{Effect of increasing the contrast between the two values of the noise in the interface profile. The values used are: (a) $\eta_{A}=0.3$ and  (b) $\eta_{A}=0.5$. The main difference is the switch from super-rough scaling to intrinsic anomalous one.}\label{fig:intA}
\end{figure*}%

Comparing to the experimental results, our $\alpha=1.5$ differs from the value $\alpha=2.0$ obtained in Ref. \cite{SO05}. However, the other independent exponents,  $\alpha_{loc}$ and $z$ are in good agreement. This discrepancy in the $\alpha$ exponent could be attributed to  a very low velocity of the experimental interface, not attained in the numerical integration of the phase field, introducing a different propertie on the scaling of the quenched noise; however, it remains as an open question.
\subsection{Variation in the capillary contrast}\label{SubSecIVB}
Another interesting issue to study is what happens when we increase the contrast between the two values of the noise keeping constant the applied pressure. Actually, it is similar to increase the strongness of the capillarity of the Hele-Shaw cell and a similar study was carried out experimentally by Soriano \emph{et al} \cite{SO03}. So in this sense, we are interested to study the change of the scaling exponents when we increase the parameter $\eta_{A}$. Our starting point is the known results of super-rough interfaces with $\alpha=1.25$ and $z=4$ obtained in the case of $\mu_{0}=-1.0$ and $\eta_{A}=0.3$. In Fig.  \ref{fig:intA}  we can see two different interface profile with two different capillary contrast, $\eta_{A}=0.3$ and $\eta_{A}=0.5$ respectively, where we observe the morphological difference between both interfaces. While in the former case the interfaces are characterized by a smooth profile,  large local slopes appear when the capillary contrast is increased. Note that due to numerical limitations, the capillary contrast is not able to be too large. Actually, for values $\eta_{A}>0.6$ we have observed that the interface becomes unstable and a clear statistical description of the interface fluctuations can not be performed.

\begin{figure}[!]%
 \centering %
 \subfiguretopcapfalse
 \subfigure{{\includegraphics[width=0.45\textwidth,origin=c,angle=0]{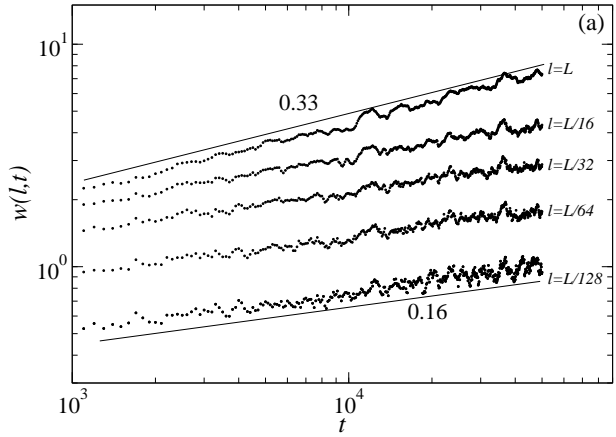}}}\\%
 \vspace*{0.6cm}%
 \subfigure{\includegraphics[width=0.45\textwidth,origin=c,angle=0]{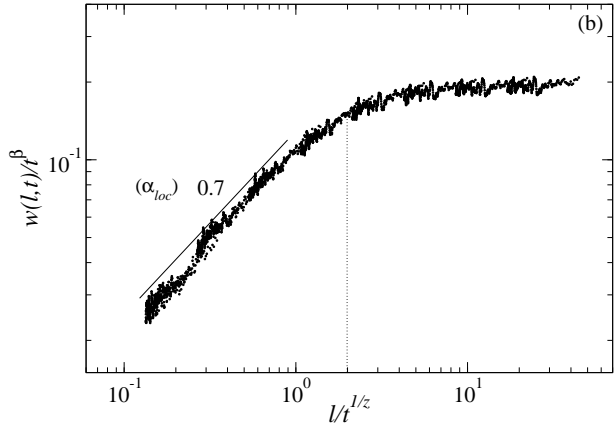}}\\%
 \vspace*{0.6cm}%
 \subfigure{\includegraphics[width=0.45\textwidth,origin=c,angle=0]{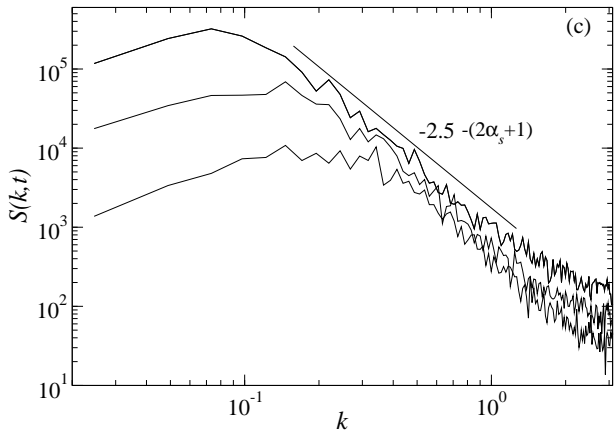}}%
 \caption{Statistical description of interface with an increased capillary contrast, $\eta_{A}=0.5$: (a) Log-log plot of the interfacial width as function of time in different windows sizes $\ell$. The data fit shows a growth exponent $\beta= 0.33\pm 0.03$ and and the local one $\beta^{*}= 0.16\pm 0.04$.(b) Collapse of the interface local width function, $w(\ell,t)$, which is scaled by $t^{\beta}$ versus $\ell /t^{\beta/\alpha}$, where $\alpha=1.33\pm 0.03$, and then $z=4.0\pm 0.1$  follows from $z=\alpha/\beta$. The picture also shows a local roughness exponent $\alpha_{loc}= 0.70 \pm 0.03$. (c) Power spectra of the interface $h(x,t)$ at different times. It shows that $\alpha_{s}=0.75\pm 0.05$.}\label{fig:StatisticalAZ4}
\end{figure}

The different scaling exponents for the case of $\eta_{A}=0.5$ are computed in Fig. \ref{fig:StatisticalAZ4} obtaining: $\alpha=1.33 \pm 0.03$, $z=4.0 \pm 0.1$, $\beta = 0.33\pm 0.03$, $\beta^{*}= 0.16\pm 0.04$ and $\alpha_{loc}= 0.70 \pm 0.03$, which are compatible within the intrinsic anomalous scaling framework. In Table \ref{tab:expA} we can realize how the scaling exponents change by varying the noise contrast. While the global exponents ($\alpha$, $\beta$ and $z$) are quite similar in both cases, the interface local behavior changes from a super-rough to an intrinsic anomalous description. 

If we use a gaussian noise instead of a dichotomic noise as in the previous section, we obtain a similar variation in the local scaling properties of the interface fluctuations. It turns  out that increasing the strength of the gaussian noise, $\Delta\eta$, an intrinsic anomalous scaling of the interface fluctuations is observed. Therefore, there seems to be a relation between the capillary contrast of the dichotomic noise and the strength of the gaussian field. Likewise, Laurila \emph{et al} \cite{LA05} have observed a slightly dependence of the strength of gaussian noise on the roughness exponent $\alpha$ in a similar way that our results of Table \ref{tab:expA} when we change the capillary contrast of the dichotomic noise.
\begin{table}[h]%
\centering%
\begin{ruledtabular}
\begin{tabular}{c|cccccc}%
  $\eta_{A}$ &  $\alpha$  &  $z$  & $\beta$ & $\beta^{*}$   &  $\alpha_{s}$  &  Scaling\\ \hline%
  0.3   &     1.25   &    4   &   0.32  &    0.10  &    1.25    &   S-R  \\
  0.5  &     1.33   &    4   &   0.33  &    0.16  &     0.65     &   I-A   \\
 \end{tabular}%
\end{ruledtabular}
\caption{Scaling exponents for two different capillary contrast, $\eta_{A}$,  with an applied pressure $\mu_{a}=-1.0$.}\label{tab:expA}%
\end{table}%
\subsubsection*{Scaling properties of the quenched noise}
\begin{figure}[!]%
 \centering %
 \includegraphics[width=0.4\textwidth,origin=c]{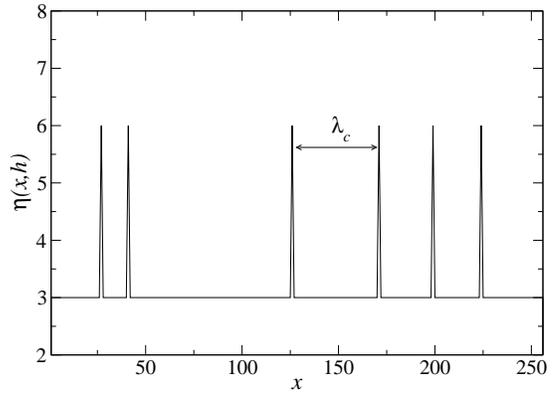}
 \caption{Dichotomic noise, $\eta(x,h)$, evaluated in a fixed interface $h(x,t_{f})$ for each $x$ in the case of the capillary contrast $\eta_{A}=0.5$. $\lambda_{c}$ represents an averaged length where the noise takes a constant value. }\label{fig:NoiseT}
\end{figure}%
Our objective now is to try to determine analytically the global scaling exponents observed in each case of the studied capillary contrast. While the dynamic exponent $z$ can be well understood from the different terms of the linear equation, Eq. (\ref{eq:LinealH}),  the roughness exponent $\alpha$ involves the knowledge of the scaling of the noise and it can be a more difficult task.

First, we  assume the general form for the noise correlation 
\begin{equation}\label{eq:NoiseCorr}
\langle\eta(x,h)\eta(x',h')\rangle\simeq f(x,x')g(h,h').
\end{equation}
In order to describe the scaling of the noise, we are interested in find out the expression for the functions $f(x,x')$ and $g(x,x')$. To do it, we first focus on the case of  $\eta_{A}=0.5$ (Fig. \ref{fig:IAInt}). In this situation, an important feature of the interface velocity is that it has  a great increase  when the interface cross through the sites where the noise takes the largest value of Eq. (\ref{eq:Noise}). In some way, interface seems to avoid these sites and it prefers to be where the noise is lower, that is $\eta=\eta_{0}$.  It leads to think that there has to be a length $\lambda_{c}$ below which the noise has a constant value. Since there is no imposed velocity but the interface moves only due to the capillarity of the media, this length $\lambda_{c}(t,\eta_{A})$ is expected to increase with time and with the capillary contrast. An interpretation of $\lambda_{c}$ can be seen in Fig. \ref{fig:NoiseT}, where  the values of the quenched noise $\eta(x,h)$ in a fixed interface $h(x,t_{f})$ are plotted,  being $t_{f}$  an arbitrary chosen time. Note that $\lambda_{c}$ has to be understood as an average length in such way that we can suppose the function $f(x,x')\simeq const$ for $\vert x-x'\vert<\lambda_{c}$ and $f(x,x')\simeq \delta(x-x')$ for $\vert x-x'\vert>\lambda_{c}$.

\begin{figure}[!]%
 \centering %
\includegraphics[width=0.4\textwidth,origin=c,angle=0]{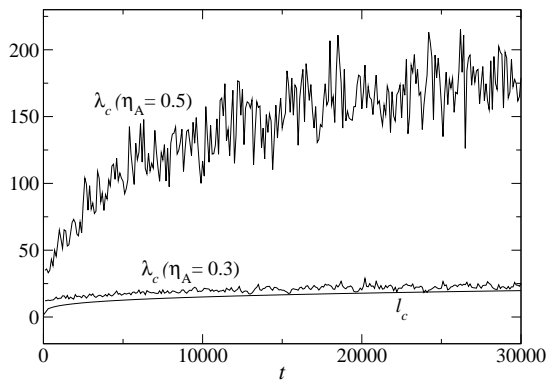}%
\caption{The length $\lambda_{c}$ of the quenched noise is always greater than interface correlation length $\ell_{c}$ in both cases of $\eta_{A}=0.3$ and $0.5$.}\label{fig:Lambda}
\end{figure}

Since the only length scale characterizing the fluctuations growth in the $x$ direction is the correlation length $\ell_{c}$, two different  scaling for the noise emerge depending if $\ell_{c}>\lambda_{c}$ or $\ell_{c}<\lambda_{c}$.  In Fig. \ref{fig:Lambda} there is plotted the length $\lambda_{c}$ in the two cases of capillary contrast, $\eta_{A}=0.3$ and $\eta_{A}=0.5$, compared with the interface correlation length $\ell_{c}\simeq t^{1/4}$. We can see that in both cases $\lambda_{c}>\ell_{c}$ and thus we can take the correlation of the noise along the $x$-direction as a constant. 

Concerning the expression for the function $g(h,h')$, we will suppose the simplest case of $g(h,h')=\delta(h-h')$. However,  note that  two different  kind of scaling emerge depending if the quenched noise is evaluated on the interface $\eta(x,h)$ or on the  mean value of the interface,   $\eta(x,h)\simeq \eta(x,H(t))$. Therefore, we conclude that the quenched noise has to scale as 
\begin{align}
\eta(x,h)\to & b^{0}b^{-\frac{\alpha}{2}}\eta(x,h), \label{eq:NoiseScalh}{} \\ 
\eta(x,H)\to & b^{0}b^{-\frac{z}{4}}\eta(x,H),\label{eq:NoiseScalH}
\end{align}
when we rescale it by Eq. (\ref{eq:rescaling}). The term with $b^{0}$ comes from the constant function $f(x,x')$. We have also assumed the Washburn law $H(t)\simeq t^{1/2}$.

It should be noted that these noise scalings are not valid when the interface movement is near a critical pinning situation ( as in the previous section where $\mu_{a}\sim -3.0$) since in such a case there is an important number of sites where the interface is locally pinned and the effects of a columnar noise, Eq. (\ref{eq:ScaleColumnarNoise}), become important.
\subsubsection*{Global scaling exponents}
Once we know how the noise scales,  we can compute analytically  the global scaling exponents in the studied cases of capillary contrast, $\eta_{A}=0.3$ and $\eta_{A}=0.5$. We do a scaling analysis of the linear equation, Eq. (\ref{eq:LinealH}), taking into account only the surface tension term. Imposing the noise scaling as Eq. (\ref{eq:NoiseScalh}), we obtain the exponent $\alpha=1.33$, which is in accordance to the case of $\eta_{A}=0.5$. On the other hand, if we suppose that noise scales as Eq. (\ref{eq:NoiseScalH}), the exponent is then $\alpha=1.25$, which adjusts the numerical value obtained in Ref. \cite{DU99} and in our case of $\eta_{A}=0.3$. Note that the $\alpha=1.33$  observed when $\eta_{A}=0.5$ is the same that the value obtained in the Sec. \ref{SubSecIVA}, in the case of $\mu_{a}=-2.0$ (see Table \ref{tab:exp}), indicating that, in that case,  the scaling of the noise has to follows also Eq. (\ref{eq:NoiseScalh}). Therefore, we analytically obtain the roughness exponents in the high velocity regime when we vary either the capillary contrast or the applied pressure.

Eventually, since the crossover length controls the interface fluctuations in both cases, we conclude that the dynamic exponent has to be $z=4$, and then we achieve the whole set of global scaling exponents described in Table \ref{tab:expA}.
\section{Conclusions}\label{SecV}
We have presented numerical results of  spontaneous imbibition using a phase field model. Our study has been focused on the dependence of the scaling exponents when we vary either the applied pressure imposed at the origin of the system or the capillary contrast of the dichotomic quenched noise.

By decreasing the applied pressure, we have got interfaces advancing at low velocities. In this situation of spontaneous imbibition, our results are in good agreement with the experimental work made in Ref. \cite{SO05}. Indeed, our numerical study confirms that there is a regime of low velocities where the growth of the interface fluctuations is controlled by the surface tension of the interface, following $\ell_{c}\simeq t^{1/3}$. We have also observed that the \emph{cut off} crossover length, $\xi_{\times}$, acts as an effective correlation length when the initial interface velocity is higher, accordingly to the results of Dub\'e \emph{et al} \cite{DU99}.

On the other hand, our results show that the local scaling properties of the interface change as we decrease the applied pressure. More precisely, we obtain an intrinsic anomalous description at the regime of low velocities, where the interface is locally pinned. In addition, the roughness exponent we observe in such a situation is $\alpha=1.5$ which is the same value obtained in the local QEW model with a columnar noise, $\eta(x)$.
 
It is also important to remark that  the intrinsic anomalous scaling is not  related to the low velocity of the interface but to the capillary contrast of the noise. Indeed, when we increase the contrast between both noise values keeping constant the applied pressure, the local fluctuations  description of the interface changes from a super-rough to an intrinsic anomalous scaling, indicating the existence of a direct relation of the capillary contrast and the intrinsic anomalous scaling, which had been observed experimentally in Ref. \cite{SO03}. We have also introduced some scaling properties of the dichotomic quenched noise which have allowed us to determine analytically the obtained numerical scaling exponents.
\begin{acknowledgments}
Fruitful discussions with  J.M. L\'opez, M.A. Rodr\'iguez, J. Ort\'in , R. Planet and A.M. Lacasta are acknowledged. The research has received financial support from Project BFM2003-07749-C05. M.P. is supported by a fellowship of the MEC, Spain. 
\end{acknowledgments}


\begin{thebibliography}{99}
\bibitem{AL04}{M. Alava, M. Dub\'e and M. Rost, \emph{Adv. Phys.} \pmb{53}, 83 (2004).}
\bibitem{AN04}{T. Ala-Nissila, S. Majaniemi and K. Elder, \emph{Lect. Notes Phys.} \pmb{640}, 357 (2004).}
\bibitem{BO97}{E. Bouchard, \emph{J. Phys. Cond. Matt.} \pmb{9}, 4319 (1997).}
\bibitem{LO98}{J.M. L\'opez and J. Schimttbuhl, \emph{Phys. Rev. E} \pmb{57}, 6405 (1998).}
\bibitem{MA97}{J. Maunuksela, M. Myllys, O.-P. K\"ahk\"onen, J. Timonen, N. Provatas, M.J. Alava and T. Ala-Nissila, \emph{Phys. Rev. Lett.} \pmb{79}, 1515 (1997).}
\bibitem{SO05}{J. Soriano, A. Mercier, R. Planet, A. Hern\' andez-Machado, M.A. Rodr\'iguez and J. Ort\'in, \emph{Phys. Rev. Lett.} \pmb{95}, 104501 (2005).}
\bibitem{SO02}{J. Soriano, A. Hern\' andez-Machado and J. Ort\'in, \emph{Phys. Rev. E} \pmb{66}, 031603 (2002).}
\bibitem{SO02b}{J. Soriano, J.J. Ramasco, M.A. Rodr\'iguez, A. Hern\' andez-Machado and J. Ort\'in, \emph{Phys. Rev. Lett.} \pmb{89}, 026102 (2002).}
\bibitem{SO03}{J. Soriano, A. Hern\' andez-Machado and J. Ort\'in, \emph{Phys. Rev. E} \pmb{67}, 056308 (2003).}
\bibitem{GE02}{D. Geromichalos, F. Mugele and S. Herminghaus, \emph{Phys. Rev. Lett} \pmb{89}, 104503 (2002).}
\bibitem{BA95}{A.-L. Barab\' asi, H.E. Stanley, \emph{Fractal Concepts in Surface Growth} (Cambridge University Press, Cambridge,  England, 1995).}
\bibitem{FA85}{F. Family and T. Vicsek, \emph{J. Phys. A} \pmb{18}, L75 (1985).}
\bibitem{LE93}{H. Leschhorn and L.-H. Tang, \emph{Phys. Rev. Lett.} \pmb{70}, 2973 (1993).}
\bibitem{LO97}{J.M. L\'opez, M.A. Rodr\'iguez and R. Cuerno, \emph{Physica A} \pmb{246}, 329 (1997).}
\bibitem{LO96}{J.M. L\'opez and M.A. Rodr\'iguez, \emph{Phys. Rev. E} \pmb{54}, R2189 (1996).}
\bibitem{LO99}{J.M. L\'opez, \emph{Phys. Rev. Lett.} \pmb{83}, 4594 (1999).}
\bibitem{RA00}{J.J. Ramasco, J.M. L\'opez and M.A. Rodr\'iguez, \emph{Phys. Rev. Lett} \pmb{84}, 2199 (2000).}
\bibitem{HM01}{A. Hern\'andez-Machado, J. Soriano, A.M. Lacasta, M.A. Rodr\'iguez, L. Ram\'irez-Piscina and J. Ort\'in, \emph{Europhys. Lett.} \pmb{55}, 194 (2001).}
\bibitem{PA03}{E. Paun\'e and J. Casademunt, \emph{Phys. Rev. Lett.} \pmb{90}, 144504 (2003).}
\bibitem{DU99}{M. Dub\'e, M. Rost, K.R. Elder, M. Alava, S. Majaniemi and T. Ala-Nissila, \emph{Phys. Rev. Lett.} \pmb{83}, 1628 (1999).}
\bibitem{DU00}{M. Dub\'e, M. Rost and M. Alava, \emph{Eur. Phys. J. B} \pmb{15}, 691 (2000); M. Dub\'e, M. Rost and M. Alava, \emph{Eur. Phys. J. B} \pmb{15}, 701 (2000).}
\bibitem{EL01}{K.R. Elder, M. Grant, N. Provatas and J.M. Kosterlitz, \emph{Phys. Rev. E} \pmb{64}, 021604 (2001).}
\bibitem{LA05}{T. Laurila, C. Tong, I. Huopaniemi, S. Majaniemi and T. Ala-Nissila, \emph{Eur. Phys. J. B} \pmb{46}, 553 (2005).}
\bibitem{LU05}{K. Luo, M.-P Kuittu, C. Tong, S. Majaniemi and T.Ala-Nissila, \emph{J. Chem. Phys.} \pmb{123}, 194702 (2005).}
\bibitem{LA06}{T. Laurila, C. Tong, I. Huopaniemi, S. Majaniemi and T. Ala-Nissila, \emph{cond-matt}/0601473 (2006).}
\bibitem{WA21}{E.W. Washburn, \emph{Phys. Rev.} \pmb{17}, 273 (1921).}
\bibitem{HH77}{P.C. Hohenberg and B.I. Halperin, \emph{Rev. Mod. Phys.} \pmb{49}, 435 (1977).}
\bibitem{HM03}{A. Hern\'andez-Machado, A.M. Lacasta, E. Mayoral and E. Corvera-Poir\'e, \emph{Phys. Rev. E}, \pmb{68}, 046310 (2003).}
\bibitem{PA92}{G. Parisi, \emph{Europhys. Lett.} \pmb{17}, 673 (1992).}
\bibitem{LO95}{J.M. L\' opez and M.A. Rodr\'iguez, \emph{Phys. Rev. E} \pmb{52}, 6442 (1995).}
\end{thebibliography}
\end{document}